\documentclass[draft, a4paper]{article}
\usepackage[english]{babel}
\usepackage{amsmath,amssymb,amsfonts,amstext}
\usepackage{url}
\usepackage{float}

\usepackage{bm}
\renewcommand{\v}[1]{\bm{ #1 }}

\usepackage{geometry}
\begin{document}

\markboth{L.~De~Cruz, D.~G.~Ireland, P.~Vancraeyveld, J.~Ryckebusch}
{Bayesian Model Selection for Electromagnetic Kaon Production on the Nucleon}

\title{BAYESIAN MODEL SELECTION FOR ELECTROMAGNETIC KAON PRODUCTION ON THE NUCLEON}

\author{L.~De~Cruz$^{\circ}$$^,$\footnote{E-mail address: Lesley.DeCruz@UGent.be}~,
D.~G.~Ireland$^{\&}$, 
P.~Vancraeyveld$^{\circ}$,
J.~Ryckebusch$^{\circ}$\\ \\
\emph{$^{\circ}$ Department of Physics and Astronomy, Ghent University,}\\ 
\emph{Proeftuinstraat 86, B-9000 Gent, Belgium}\\
\emph{$^{\&}$Department of Physics and Astronomy, University of Glasgow,}\\ 
\emph{Glasgow G12 8QQ, United Kingdom }
}

\maketitle

\begin{abstract}	
We present the results of a Bayesian analysis of a Regge model for $K^+\Lambda$ photoproduction. The model is based on the exchange of $K^{+}(494)$ and $K^{\ast+}(892)$ trajectories in the $t$-channel. For different prior widths, we find decisive Bayesian evidence ($\Delta \ln \mathcal{Z} \approx $ 24) for a $K^+\Lambda$ photoproduction Regge model with a positive vector coupling and a negative tensor coupling constant for the $K^{\ast+}(892)$ trajectory, and a rotating phase factor for both trajectories. Using the $\chi^2$ minimization method, one could not draw this conclusion from the same dataset.  \\
\emph{Keywords:}  Regge phenomenology; Bayesian inference; Model selection\\
PACS numbers: 11.55.Jy, 12.40.Nn, 13.60.Le
\end{abstract}

% 11.10.Ef 	Lagrangian and Hamiltonian approach 
% 11.55.Jy 	Regge formalism
% 12.40.Nn 	Regge theory, duality, absorptive/optical models
% 13.60.Le 	Meson production 
% 14.20.Gk 	(properties of) Baryon resonances with S=0 

\section{Introduction}

The study of electromagnetic open strangeness or kaon-hyperon ($KY$) production is a key step towards understanding the structure of the nucleon\cite{capstick-2000,saghai-2006}. The focus of such studies has recently shifted from parameter estimation to model comparison, in part due to the importance of identifying the set of contributing resonances\cite{adelseck-1985,adelseck-1988,corthals-2006}. The statistical tools, however, have not been adapted to this new objective. The least-squares method in particular has often been stretched beyond its limits, being used not only as an optimization tool, but also as a model selection criterion. We advocate the Bayesian evidence as a more robust and well-founded tool for model comparison, and apply it to a Regge model for $K^+\Lambda$ photoproduction.

\section{Bayesian analysis}\label{sec:Bayes}

The quantity we will use to quantify model fitness is the Bayesian evidence or marginal likelihood $\mathcal{Z}$, which is proportional to the posterior probability $P(M|\left\lbrace d_k \right\rbrace)$ of a model $M$ given data $\left\lbrace d_k \right\rbrace$\cite{ireland-2008}.  
$\mathcal{Z}$ can be calculated by integrating over the model parameters $\v{\alpha}_M$
\begin{align}
\mathcal{Z} \equiv P(\left\lbrace d_k \right\rbrace|M)
		&= \int \underbrace{P(\left\lbrace d_k \right\rbrace|\v{\alpha_M},M)}_{ \mathcal{L}(\v{\alpha_M})} \, \underbrace{P(\v{\alpha_M}|M)}_{\pi(\v{\alpha_M})}\, d\v{\alpha_M}.
\end{align}
We employ a uniform prior $\pi(\v{\alpha_M})$ and use the following limiting distribution to approximate the likelihood function\cite{barlow-1989}
\begin{equation}
 \mathcal{L}(\v{\alpha_M}) \approx \frac{1}{2\sqrt{\pi k}}\exp{-\frac{(\chi^2(\v{\alpha_M}) - k)^2}{4k}}.
\end{equation}
To compute the Bayesian evidence integrals, we employ the Nested Sampling (NS) method, developed by Skilling\cite{skilling-2006}. The results can be interpreted qualitatively using Jeffreys' scale\cite{jeffreys-1961}, i.e. $\Delta\ln{\mathcal{Z}} > 5$ decisively favors the model with the highest evidence.

\section{Bayesian analysis of a Regge Model} \label{sec:Regge}

The Regge model under investigation is based on the exchange of the $K^{+}(494)$ and $K^{\ast+}(892)$ trajectories\cite{vanderhaeghen-1997,guidal-2003}. The amplitude is derived from the $t$-channel Feynman amplitude by replacing the Feynman propagator by the respective Regge propagator, as described in Ref.~\cite{corthals-2006}. The model's free parameters are the so-called sign factors (either 1, constant phase or $e^{-i\pi\alpha(t)}$, rotating phase) and three continuous parameters, namely the strong coupling constant $g_{K^+\Lambda p}$ of the $K^{+}$ trajectory and the tensor and vector couplings of the $K^{\ast +}$ trajectory, $G_{K^{\ast +}}^{v,t} = e \, g_{{K^{\ast +}}\, \Lambda p}^{v,t}\ \kappa_{K^+{K^{\ast +}}}\;/4\pi$.

Optimization of these parameters against the 72 available high-energy (E$_{\gamma}\gtrsim 5$ GeV) data points\cite{boyarski-1969,quinn-1979,vogel-1972} reveals that there are several model variants with comparable $\chi^2$ values\cite{corthals-2006}. The sign factors ($e^{-i\pi\alpha(t)}$ or 1) as well as the signs of $G^{v}_{K^{\ast+}}$ and $G^{t}_{K^{\ast+}}$ cannot be established conclusively using the $\chi^2$-method. These sign and phase ambiguities may not seem important for the Regge model itself. However, for the Regge-plus-resonance (RPR) model, in which the Regge background is complemented with $s$-channel nucleon ($N^{\ast}$) resonances, an exact determination of the background parameters is of major importance as it affects the extraction of the resonance information. 

\section{Results}\label{sec:Results}

We employ a uniform prior for all continuous parameters and allow for a 20\% deviation from the SU(3) prediction for the coupling constant $g_{K^+\Lambda p}$ when determining its prior interval\cite{guidal-phd}. For $G_{K^{\ast +}}^{v,t}$, we compare results using prior widths of 100, 1000 and 10000 to ensure that the results do not depend on the choice of the prior. Indeed, multiplying the prior width with a factor of 10 results in a difference of less than $5\%$ in the computed values of $\ln{\mathcal{Z}}$. Thereby, one should bear in mind that a reduced sampling efficiency leads to considerably larger errors. Repeating the calculations assuming 40\% SU(3) symmetry breaking does not significantly influence the results. More importantly, the ranking of the models is not affected by the above prior modifications\cite{decruz-2010}.

We find that the value of $\ln\mathcal{Z}$ for the best model variant for $K^{+}\Lambda$ photoproduction, with a rotating phase for both $K^{+}$ and $K^{\ast+}$ trajectories, positive vector and negative tensor coupling, is $24.30$ $\pm$ $0.75$ above the second-best variant. This result decisively resolves the sign and phase ambiguity for $K^{+}\Lambda$ photoproduction, which could not be achieved in a previous fit of the model to the high-energy data using the $\chi^2$ method\cite{corthals-2006}.

\section{Conclusions and outlook}\label{sec:conclusion}

Bayesian inference provides us with an excellent tool for model comparison. We have demonstrated this by using the Nested Sampling algorithm to compute the Bayesian evidence for different model variants of a Regge model for $K^{+}\Lambda$ photoproduction. 
The Nested Sampling method has many applications, both for the RPR model and for other research. One of these applications is the accurate estimation of model parameters as well as the elimination of nuisance parameters. More importantly, however, this method may provide us with a means to address the missing-resonance problem by calculating the probability of individual resonance contributions in a Bayesian framework. This is an approach we intend to explore in the near future.

\section*{Acknowledgements}

This research was funded by the Research Foundation - Flanders (FWO Vlaanderen). D.G.I. acknowledges the support of the UK Science and Technology Facilities Council.

\end{document}